# Minimum information and guidelines for reporting a Multiplexed Assay of Variant Effect

Melina Claussnitzer[1,2,†], Victoria N. Parikh[3,†], Alex H. Wagner[4,5,6,†], Jeremy A. Arbesfeld[4], Carol J. Bult[7], Helen V. Firth[8,9], Lara A. Muffley[10], Alex N. Nguyen Ba[11], Kevin Riehle[12], Frederick P. Roth[13,14,15,16], Daniel Tabet[13,14,15,16], Benedetta Bolognesi[17,*], Andrew M. Glazer[18,*], Alan F. Rubin[19,20,*]

1 The Novo Nordisk Foundation Center for Genomic Mechanisms of Disease, Broad Institute of MIT and Harvard, Cambridge, MA 02142, USA
2 Center for Genomic Medicine, Massachusetts General Hospital, Harvard Medical School, Cambridge, MA 02142, USA
3 Stanford Center for Inherited Cardiovascular Disease, Stanford University School of Medicine, Stanford, CA, USA 94305
4 The Steve and Cindy Rasmussen Institute for Genomic Medicine, Nationwide Children's Hospital, Columbus, OH 43215, USA
5 Department of Pediatrics, The Ohio State University College of Medicine, Columbus, OH 43210, USA
6 Department of Biomedical Informatics, The Ohio State University, Columbus, OH 43210, USA
7 The Jackson Laboratory, Bar Harbor, ME 04609, USA
8 Wellcome Sanger Institute, Hinxton, Cambridge, UK
9 Dept of Medical Genetics, Cambridge University Hospitals NHS Trust, Cambridge UK
10 Department of Genome Sciences, University of Washington, Seattle, WA 98105, USA
11 Department of Biology, University of Toronto at Mississauga, Mississauga, Ontario, Canada
12 Department of Molecular and Human Genetics, Baylor College of Medicine, Houston, TX 77030, USA
13 Donnelly Centre, University of Toronto, Toronto, Ontario, Canada
14 Department of Molecular Genetics, University of Toronto, Toronto, Ontario, Canada
15 Department of Computer Science, University of Toronto, Toronto, Ontario, Canada
16 Lunenfeld-Tanenbaum Research Institute, Sinai Health, Toronto, Ontario, Canada
17 Institute for Bioengineering of Catalunya (IBEC), The Barcelona Institute of Science and Technology, Barcelona, Spain
18 Vanderbilt University Medical Center, Nashville, TN 37232, USA
19 Bioinformatics Division, WEHI, Parkville, Victoria, Australia
20 Department of Medical Biology, University of Melbourne, Parkville, Victoria, Australia
† These authors contributed equally to this work.
* To whom correspondence should be addressed bbolognesi@ibecbarcelona.eu, andrew.m.glazer@vumc.org, alan.rubin@wehi.edu.au.

## Abstract

Multiplexed Assays of Variant Effect (MAVEs) have emerged as a powerful approach for interrogating thousands of genetic variants in a single experiment. The flexibility and widespread adoption of these techniques across diverse disciplines has led to a heterogeneous mix of data formats and descriptions, which complicates the downstream use of the resulting datasets. To address these issues and promote reproducibility and reuse of MAVE data, we define a set of minimum information standards for MAVE data and metadata and outline a controlled vocabulary aligned with established biomedical ontologies for describing these experimental designs.

## Keywords

# Background

The emergence of high-throughput genomic technologies has revolutionized our ability to study the impact of genetic variants at a grand scale. A prominent example of these innovative methods is Multiplexed Assays of Variant Effect (MAVEs). MAVEs are a family of experimental methods combining saturation mutagenesis with a multiplexed assay to interrogate the effects of thousands of genetic variants in a given functional element in parallel [1,2]. The output of a MAVE is a variant effect map quantifying the consequences of all single nucleotide (or single amino acid) variants in a target functional element, even variants not yet observed in the population. MAVEs have been applied to coding sequences as well as noncoding elements like splice sites and regulatory regions across various organisms. Variant effect maps have broad applications including clinical variant interpretation [2,3], understanding sequence/structure/function relationships [4,5], and investigating molecular mechanisms of evolution [6,7]. The MAVE field is growing rapidly, leading to the formation of organizations such as the Atlas of Variant Effects (AVE). AVE consists of over 400 researchers from over 30 countries who perform, interpret, and apply MAVE experiments.

The rapid growth and adoption of MAVE technologies across many fields has led to an excess of overlapping definitions, complicating discovery and interpretation. Minimum information standards in other research areas have increased the reporting, archiving, and reuse of biological data [8–11]. To promote reuse and FAIR data sharing [12], minimum information standards and a controlled vocabulary for describing MAVE experiments and variant effect maps are needed. Here, we—members of the AVE Experimental Technology and Standards and Data Coordination and Dissemination workstreams—provide a comprehensive structured vocabulary and recommendations for data release for MAVE datasets. Uptake of these recommendations by the MAVE community will greatly improve the usability and longevity of MAVE datasets, enabling novel insights and applications.

# Results and Discussion

All MAVEs share a core pipeline: generation of a variant library, delivery of the library into a model system, separation of variants based on function, quantification of variant frequency by high-throughput DNA sequencing, and carrying out data analysis and score calculation [1,2,13]. Accurate and consistent metadata describing each of these steps is the basis for the interpretability of MAVE functional scores and is a requirement for any advanced quantitative analysis, such as comparing and combining scores. To systematize these metadata, we have defined and implemented a computable controlled vocabulary that covers the majority of current and emerging MAVE techniques (**Figure 1**) [14]. This vocabulary captures the major steps of the MAVE experimental process including project scope, library generation, library integration/expression, assay type, and sequencing method. The vocabulary also contains terms to describe the biological and disease relevance of the assay. In addition to releasing scores and other datasets in published papers, we recommend sharing MAVE datasets through MaveDB, an open-source platform to distribute and interpret MAVE data [15,16].

Researchers should communicate the target sequence, the method used to generate library diversity, and the method of variant delivery into the assay system using terms from the controlled vocabulary. Metadata about the variant generation method should include terms for either editing at the endogenous locus or *in vitro* variant library generation. It should also specify the model system as defined by NCBI Taxonomy ID [17] and Cell Line Ontology (CLO) [18] terms where available.

It is essential for the target sequence to be linked to a reference genome database or similar by including a versioned stable identifier from a widely-used resource such as RefSeq [19], Ensembl [20], or UniProt [21]. We also recommend that researchers designing a study choose a reference-identical allele when it does not otherwise affect the study design, particularly for clinically-relevant targets. The entire target sequence used in the assay must be provided to allow MaveDB and other systems to generate globally unique identifiers (sha512t24u computed identifiers [22]) as used by the Global Alliance for Genomics and Health (GA4GH) [23] refget [24] and Variation Representation Specification (VRS) [25] standards.

We recommend that variant libraries are exchanged using VRS and stored using a VRS-compatible information model, including the aforementioned computed identifiers, inter-residue sequence location data, and VOCA-normalized allele representation [25,26]. This allows variants to be defined in terms of both the variant on the target sequence and the homologous variant on the linked reference sequences with an appropriate variant mapping relation, such as the *homologous_to* relation from the sequence ontology [27]. Descriptions of variants on target sequences should follow the MAVE-HGVS nomenclature conventions [16]. Homologous variants on linked reference sequences should describe variants following conventions typical for the target organism, e.g. using the Human Genome Variation Society (HGVS) variant nomenclature [28] for variants on human reference sequences. An example of these sequence variant recommendations in practice are described by Arbesfeld et al. [29], where they enable interoperability with downstream resources including the Ensembl Variant Effect Predictor (VEP) [30], UCSC Genome Browser [31], the Genomics to Proteins resource [32], the ClinGen Allele Registry [33] and ClinGen Linked Data Hub.

The phenotypic assay is the most unique aspect of a MAVE compared to other data types for which minimum information standards have been established. There is a tremendous diversity in functional assays in terms of both the assay readout and the biology the assay was designed to interrogate. For assay readout, we have identified a subset of phenotypic readouts in the Ontology for Biomedical Investigations (OBI) [34] that are commonly used in variant effect maps. Because OBI has over 2,500 terms, we hope that this "short list" will help researchers identify the most relevant terms to describe their experiments. Neverthelesswe also welcome the use of other OBI terms if necessary to describe new assays. Researchers should also detail any environmental variables (such as the addition of small molecules) and use the appropriate controlled vocabulary term for the high-throughput sequencing method used for variant quantification. We strongly recommend that raw sequence reads be deposited in a suitable repository, such as the Sequence Read

Archive (SRA) [35] or Gene Expression Omnibus (GEO) [36], along with a description of each file (e.g. time point and sample information).

We recommend that researchers investigating clinical phenotypes use terms from the Mondo Disease Ontology (Mondo) [37] or Online Mendelian Inheritance in Man (OMIM) [38] to help clinicians and other stakeholders retrieve relevant functional data. Particular care is needed for genes encoding proteins with multiple functional domains and where loss of function and gain of function variants are associated with different diseases. Ideally, each MAVE should be associated with a particular gene-disease entity that describes the mechanism of disease such as those defined by G2P [39] and how the MAVE assay recapitulates or is relevant to the mechanism of disease. Some genes or functional domains may require multiple MAVE assays, each probing a different function or attribute of the gene product, to accurately model different disease entities.

Although it is not within the scope of this controlled vocabulary, it is still crucial to detail the data analysis performed to produce a variant effect map. This includes steps to generate variant counts, including sequence read processing, quality filtering, alignment, and variant identification, as well as further statistical and bioinformatic processing to calculate scores and associated error estimates. Researchers should describe the analysis pipeline used for these calculations, including software versions. Several well-documented tools are available for this purpose and the field continues to advance rapidly [40–42]. Custom code should be shared using GitHub or a similar platform and archived using Zenodo or a similar archival service that mints a DOI.

In addition to processed variant scores, we urge researchers to share raw counts for each dataset, as these have tremendous utility for downstream users who want to reanalyze datasets or develop new statistical methods. Similarly, we recommend that researchers also report scores prior to normalization or imputation, and MaveDB supports the deposition of counts, scores, normalized/imputed scores, and sequence metadata for the same dataset (**Table 1**).

## Conclusions

Minimum data standards are important to guide researchers who want their datasets to be used and cited broadly. We anticipate that this document will enhance the readability and discoverability of current and future datasets by defining a vocabulary that can be adopted across the many fields where MAVEs are being performed and where the resulting datasets are being used. Ensuring a minimum set of available metadata that uses a shared set of terms enables new types of analysis, such as machine learning methods to combine large numbers of disparate, high-dimensional datasets like MAVEs. Large-scale meta-analyses of multiple MAVE datasets have already been implemented in several contexts, including computational prediction of variant effects [43,44] and clinical variant reclassification [45]. In the near term, the controlled vocabulary will be implemented as part of MaveDB records, creating a large set of rich metadata annotations that can be searched and mined. We believe that the MAVE community should share datasets and resources responsibly, and that accessibility is real only when it ensures usability and reproducibility.

## Methods

The initial draft of the controlled vocabulary was developed collaboratively using Google Docs. The controlled vocabulary schema is defined using JSON Schema Draft 2020-12.

## Declarations

**Ethics approval and consent to participate**
Not applicable

**Consent for publication**
Not applicable

**Availability of data and materials**
The controlled vocabulary implementation is available on GitHub from the AVE Data Coordination and Dissemination workstream repository located at https://github.com/ave-dcd/mave_vocabulary and Zenodo at https://doi.org/10.5281/zenodo.8049231 [14].

**Competing interests**
The authors declare that they have no competing interests

**Funding**
MC received funding from the Novo Nordisk Foundation (NNF21SA0072102) and NIH/NIDDK grant UM1DK126185. VNP received funding from NIH/NHLBI grants K08HL143185 and R01HL164675. AHW received funding from NIH/NHGRI grant R35HG011949. LAM, FPR, and AFR received funding from NIH/NHGRI grants UM1HG011969 and RM1HG010461. ANNB acknowledges funding from CIHR, NSERC, and the University of Toronto. KR receives funding from NIH/NHGRI grant U24HG009649. FPR received funding from a CIHR Foundation Grant. BB received funding from La Caixa Research Foundation grant LCF/PR/HR21/52410004 and the Spanish Ministry of Science, Innovation and Universities grants PID2021-127761OB-I00 and RYC2020-028861-I. AMG received funding from NIH/NHGRI grant R00HG010904 and NIH/NHLBI grant R01HL164675. This work was supported by the Australian government.

**Authors' contributions**
MC, VNP, CJB, ANNB, FPR, DT, AMG, and AFR conceptualized the study. AHW, FPR, and AFR developed the methodology. AHW and KR developed software. AHW performed validation. MC, AHW, BB, AMG, and AFR performed formal analysis. AHW conducted investigations. VNP and DT provided resources. VNP and ANNB curated data. MC, AHW, JAA, BB, AMG, and AFR wrote the original draft.MC, VNP, AHW, CJB, HVF, LAM, ANNB, DT, BB, AMG, and AFR reviewed and edited the manuscript. LAM, BB, AMG, and AFR supervised the team. VNP and LAM performed project administration.

**Acknowledgements**
The authors would like to thank Michael Boettcher and Melissa S. Cline for thoughtful discussion of this work and comments on the manuscript. We would also like to thank Alex Hopkins for administrative support. The images in Figure 1 were created in Biorender.

# References

1. Gasperini M, Starita L, Shendure J. The power of multiplexed functional analysis of genetic variants. Nat Protoc. 2016;11:1782–7.

2. Tabet D, Parikh V, Mali P, Roth FP, Claussnitzer M. Scalable Functional Assays for the Interpretation of Human Genetic Variation. Annu Rev Genet. 2022;56:441–65.

3. Starita LM, Ahituv N, Dunham MJ, Kitzman JO, Roth FP, Seelig G, et al. Variant Interpretation: Functional Assays to the Rescue. Am J Hum Genet. 2017;101:315–25.

4. Stein A, Fowler DM, Hartmann-Petersen R, Lindorff-Larsen K. Biophysical and Mechanistic Models for Disease-Causing Protein Variants. Trends Biochem Sci. 2019;44:575–88.

5. Kinney JB, McCandlish DM. Massively Parallel Assays and Quantitative Sequence–Function Relationships. Annu Rev Genomics Hum Genet. 2019;20:null.

6. Starr TN, Picton LK, Thornton JW. Alternative evolutionary histories in the sequence space of an ancient protein. Nature. 2017;549:409–13.

7. Gallego Romero I, Lea AJ. Leveraging massively parallel reporter assays for evolutionary questions. Genome Biol. 2023;24:26.

8. Brazma A, Hingamp P, Quackenbush J, Sherlock G, Spellman P, Stoeckert C, et al. Minimum information about a microarray experiment (MIAME)-toward standards for microarray data. Nat Genet. 2001;29:365–71.

9. Taylor CF, Paton NW, Lilley KS, Binz P-A, Julian RK, Jones AR, et al. The minimum information about a proteomics experiment (MIAPE). Nat Biotechnol. 2007;25:887–93.

10. Brazma A, Ball C, Bumgarner R, Furlanello C, Miller M, Quackenbush J, et al. MINSEQE: Minimum Information about a high-throughput Nucleotide SeQuencing Experiment - a proposal for standards in functional genomic data reporting. Zenodo, 2012. https://doi.org/10.5281/zenodo.5706412

11. Füllgrabe A, George N, Green M, Nejad P, Aronow B, Fexova SK, et al. Guidelines for reporting single-cell RNA-seq experiments. Nat Biotechnol. 2020;38:1384–6.

12. Wilkinson MD, Dumontier M, Aalbersberg IJJ, Appleton G, Axton M, Baak A, et al. The FAIR Guiding Principles for scientific data management and stewardship. Sci Data. 2016;3:160018.

13. Fowler DM, Fields S. Deep mutational scanning: a new style of protein science. Nat Methods. 2014;11:801–7.

14. Wagner AH, Rubin AF. ave-dcd/mave_vocabulary: 0.1.0 (v0.1.0). Zenodo; 2023. https://doi.org/10.5281/zenodo.8049231

15. Esposito D, Weile J, Shendure J, Starita LM, Papenfuss AT, Roth FP, et al. MaveDB: an open-source platform to distribute and interpret data from multiplexed assays of variant effect. Genome Biol. 2019;20:223.

16. Rubin AF, Min JK, Rollins NJ, Da EY, Esposito D, Harrington M, et al. MaveDB v2: a curated community database with over three million variant effects from multiplexed functional assays. 2021;2021.11.29.470445.

17. Schoch CL, Ciufo S, Domrachev M, Hotton CL, Kannan S, Khovanskaya R, et al. NCBI Taxonomy: a comprehensive update on curation, resources and tools. Database J Biol Databases Curation. 2020;2020:baaa062.

18. Sarntivijai S, Lin Y, Xiang Z, Meehan TF, Diehl AD, Vempati UD, et al. CLO: The cell line ontology. J Biomed Semant. 2014;5:37.

19. O'Leary NA, Wright MW, Brister JR, Ciufo S, Haddad D, McVeigh R, et al. Reference sequence (RefSeq) database at NCBI: current status, taxonomic expansion, and functional annotation. Nucleic Acids Res. 2016;44:D733-745.

20. Cunningham F, Allen JE, Allen J, Alvarez-Jarreta J, Amode MR, Armean IM, et al. Ensembl 2022. Nucleic Acids Res. 2022;50:D988–95.

21. UniProt Consortium. UniProt: the Universal Protein Knowledgebase in 2023. Nucleic Acids Res. 2023;51:D523–31.

22. Hart RK, Prlić A. SeqRepo: A system for managing local collections of biological sequences. PloS One. 2020;15:e0239883.

23. Rehm HL, Page AJH, Smith L, Adams JB, Alterovitz G, Babb LJ, et al. GA4GH: International policies and standards for data sharing across genomic research and healthcare. Cell Genomics. 2021;1:100029.

24. Yates AD, Adams J, Chaturvedi S, Davies RM, Laird M, Leinonen R, et al. Refget: standardized access to reference sequences. Bioinforma Oxf Engl. 2021;38:299–300.

25. Wagner AH, Babb L, Alterovitz G, Baudis M, Brush M, Cameron DL, et al. The GA4GH Variation Representation Specification: A computational framework for variation representation and federated identification. Cell Genomics. 2021;1:100027.

26. Holmes JB, Moyer E, Phan L, Maglott D, Kattman B. SPDI: data model for variants and applications at NCBI. Bioinforma Oxf Engl. 2020;36:1902–7.

27. Eilbeck K, Lewis SE, Mungall CJ, Yandell M, Stein L, Durbin R, et al. The Sequence Ontology: a tool for the unification of genome annotations. Genome Biol. 2005;6:R44.

28. den Dunnen JT, Dalgleish R, Maglott DR, Hart RK, Greenblatt MS, McGowan-Jordan J, et al. HGVS Recommendations for the Description of Sequence Variants: 2016 Update. Hum Mutat. 2016;37:564–9.

29. dcd-mapping [Internet]. ave-dcd; 2023 [cited 2023 May 30]. Available from: https://github.com/ave-dcd/dcd_mapping

30. McLaren W, Gil L, Hunt SE, Riat HS, Ritchie GRS, Thormann A, et al. The Ensembl Variant Effect Predictor. Genome Biol. 2016;17:122.

31. Nassar LR, Barber GP, Benet-Pagès A, Casper J, Clawson H, Diekhans M, et al. The UCSC Genome Browser database: 2023 update. Nucleic Acids Res. 2023;51:D1188–95.

32. Iqbal S, Pérez-Palma E, Jespersen JB, May P, Hoksza D, Heyne HO, et al. Comprehensive characterization of amino acid positions in protein structures reveals molecular effect of missense variants. Proc Natl Acad Sci U S A. 2020;117:28201–11.

33. Pawliczek P, Patel RY, Ashmore LR, Jackson AR, Bizon C, Nelson T, et al. ClinGen Allele Registry links information about genetic variants. Hum Mutat. 2018;39:1690–701.

34. Brinkman RR, Courtot M, Derom D, Fostel JM, He Y, Lord P, et al. Modeling biomedical experimental processes with OBI. J Biomed Semant. 2010;1 Suppl 1:S7.

35. Leinonen R, Sugawara H, Shumway M. The Sequence Read Archive. Nucleic Acids Res. 2011;39:D19–21.

36. Barrett T, Wilhite SE, Ledoux P, Evangelista C, Kim IF, Tomashevsky M, et al. NCBI GEO: archive for functional genomics data sets—update. Nucleic Acids Res. 2013;41:D991–5.

37. Mungall CJ, McMurry JA, Köhler S, Balhoff JP, Borromeo C, Brush M, et al. The Monarch Initiative: an integrative data and analytic platform connecting phenotypes to genotypes across species. Nucleic Acids Res. 2017;45:D712–22.

38. Hamosh A, Amberger JS, Bocchini C, Scott AF, Rasmussen SA. Online Mendelian Inheritance in Man (OMIM®): Victor McKusick's magnum opus. Am J Med Genet A. 2021;185:3259–65.

39. Thormann A, Halachev M, McLaren W, Moore DJ, Svinti V, Campbell A, et al. Flexible and scalable diagnostic filtering of genomic variants using G2P with Ensembl VEP. Nat Commun. 2019;10:2373.

40. Bloom JD. Software for the analysis and visualization of deep mutational scanning data. BMC Bioinformatics. 2015;16:168.

41. Rubin AF, Gelman H, Lucas N, Bajjalieh SM, Papenfuss AT, Speed TP, et al. A statistical framework for analyzing deep mutational scanning data. Genome Biol. 2017;18:150.

42. Faure AJ, Schmiedel JM, Baeza-Centurion P, Lehner B. DiMSum: an error model and pipeline for analyzing deep mutational scanning data and diagnosing common experimental pathologies. Genome Biol. 2020;21:207.

43. Wu Y, Li R, Sun S, Weile J, Roth FP. Improved pathogenicity prediction for rare human missense variants. Am J Hum Genet. 2021;108:1891–906.

44. Frazer J, Notin P, Dias M, Gomez A, Min JK, Brock K, et al. Disease variant prediction with deep generative models of evolutionary data. Nature. 2021;599:91–5.

45. Fayer S, Horton C, Dines JN, Rubin AF, Richardson ME, McGoldrick K, et al. Closing the gap: Systematic integration of multiplexed functional data resolves variants of uncertain significance in BRCA1, TP53, and PTEN. Am J Hum Genet. 2021;108:2248–58.

46. Seuma M, Lehner B, Bolognesi B. An atlas of amyloid aggregation: the impact of substitutions, insertions, deletions and truncations on amyloid beta fibril nucleation. Nat Commun. 2022;13:7084.

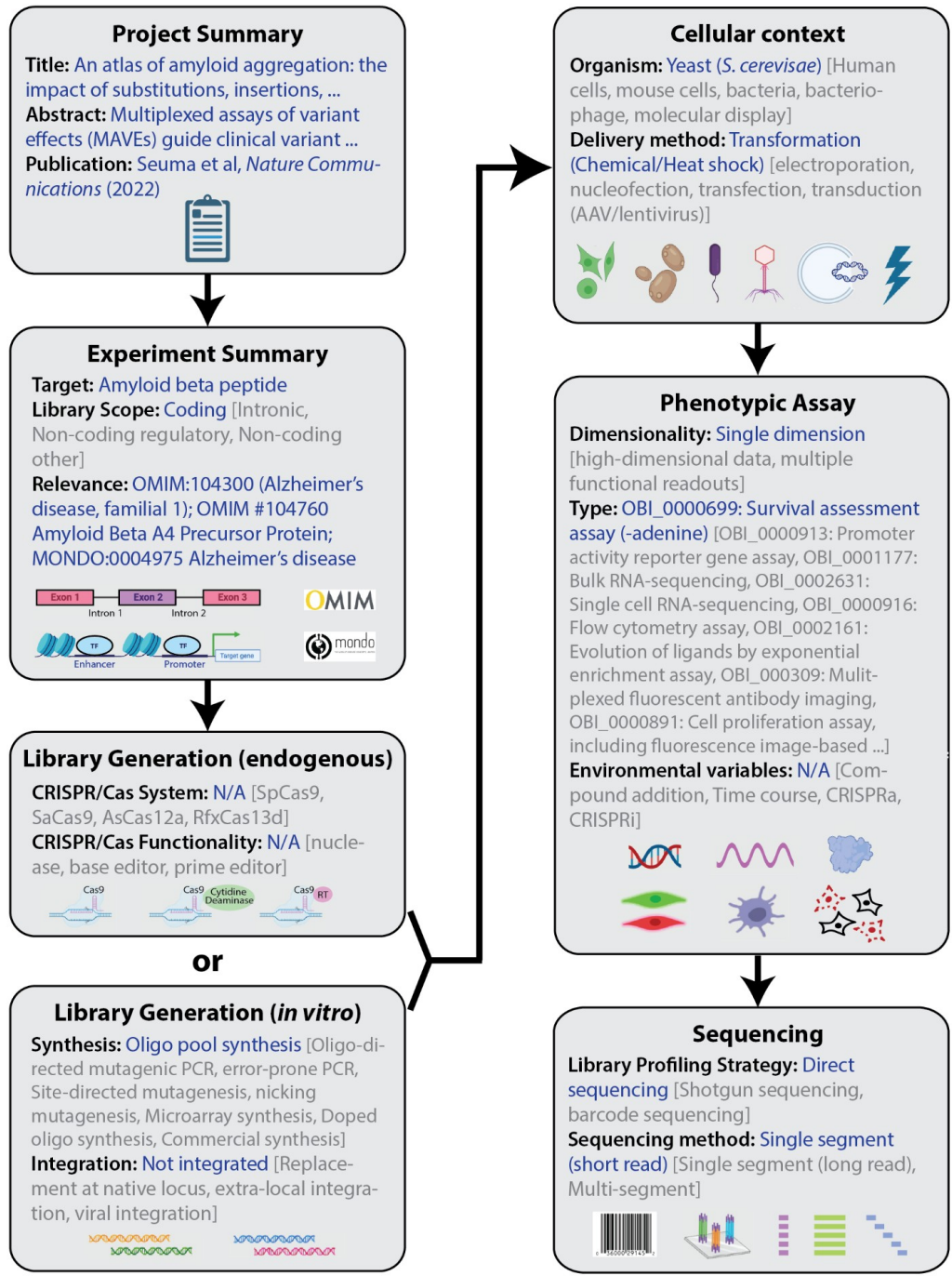

**Figure 1:** A structured vocabulary of terms relevant to the technical development, execution and recording of multiplexed assays of variant effects (MAVEs). Examples of each level of controlled vocabulary term are depicted using a published MAVE dataset [46] shown in blue with alternative options shown in gray.

**Table 1:** Recommendation locations for MAVE data deposition

| Type of data | Deposition location |
| --- | --- |
| Processed scores, unprocessed scores, raw counts | MaveDB[15,16] |
| Raw sequence reads | Sequence Read Archive[35]/Gene Expression Omnibus[36] |
| Target sequence | MaveDB[15,16] |
| Linked sequence references | MaveDB[15,16] |
| Sequence metadata / digests | MaveDB[15,16]/SeqRepo[22] |
| Variant library | MaveDB[15,16] |
| Analysis code | GitHub/Zenodo |
| Structured vocabulary description | This work/MaveDB[15,16] |